\documentclass[aps,epsfig,graphics,twocolumn,floatfix,mathbbm,a4paper]{revtex4}

\usepackage{amsmath,amsfonts,amssymb,graphics,graphicx,epsfig,color,times,bbm}

\begin{document}
\bibliographystyle{apsrev}

\newcommand{\R}{\mathbbm{R}}
\newcommand{\rr}{\mathbbm{R}}
\newcommand{\E}{{\cal E}}
\newcommand{\cc}{{\cal{C}}}
\newcommand{\ii}{\mathbbm{1}}

\newcommand{\1}{\mathbbm{1}}
\newcommand{\F}{\mathbbm{F}}

\newcommand{\tr}[1]{{\rm tr}\left[#1\right]}
\newcommand{\gr}[1]{\boldsymbol{#1}}
\newcommand{\be}{\begin{equation}}
\newcommand{\ee}{\end{equation}}
\newcommand{\bea}{\begin{eqnarray}}
\newcommand{\eea}{\end{eqnarray}}
\newcommand{\ket}[1]{|#1\rangle}
\newcommand{\bra}[1]{\langle#1|}
\newcommand{\avr}[1]{\langle#1\rangle}
\newcommand{\D}{{\cal D}}
\newcommand{\eq}[1]{Eq.~(\ref{#1})}
\newcommand{\ineq}[1]{Ineq.~(\ref{#1})}
\newcommand{\sirsection}[1]{\section{\large \sf \textbf{#1}}}
\newcommand{\sirsubsection}[1]{\subsection{\normalsize \sf \textbf{#1}}}
\newcommand{\ack}{\subsection*{\normalsize \sf \textbf{Acknowledgements}}}
\newcommand{\front}[5]{\title{\sf \textbf{\Large #1}}
\author{#2 \vspace*{.4cm}\\
\footnotesize #3}
\date{\footnotesize \sf \begin{quote}
\hspace*{.2cm}#4 \end{quote} #5} \maketitle}
\newcommand{\eg}{\emph{e.g.}~}

\newcommand{\proofend}{\hfill\fbox\\\medskip }


\newtheorem{theorem}{Theorem}
\newtheorem{proposition}{Proposition}

\newtheorem{lemma}{Lemma}

\newtheorem{definition}{Definition}
\newtheorem{corollary}{Corollary}

\newcommand{\proof}[1]{{\bf Proof.} #1 $\proofend$}

\newcommand{\alejo}[1]{{\color{red} #1}}

\title{Measurements incompatible in Quantum Theory \\ cannot be measured jointly in any other local theory}

\author{Michael M. Wolf$^1$, David Perez-Garcia$^2$, Carlos Fernandez$^3$}
\affiliation{$1$ Niels Bohr Institute, Blegdamsvej 17, 2100 Copenhagen, Denmark \\
$2$ Departamento de An\'alisis Matemático \& IMI. Universidad Complutense de Madrid. 28040 Madrid, Spain\\
$3$ Departamento de \'Algebra \& IMI. Universidad Complutense de Madrid. 28040 Madrid, Spain}
\date{\today}

\begin{abstract}It is well known that jointly measurable observables cannot lead to a violation of any Bell inequality---independent of the state and the measurements chosen at the other site. In this letter we prove the converse: every pair of incompatible quantum observables enables the violation of a Bell inequality and therefore must remain incompatible within any other no-signaling theory. While in the case of von Neumann measurements it is sufficient to use the same pair of observables at both sites, general measurements can require different choices. The main result is obtained by showing that for arbitrary dimension the CHSH inequality provides the Lagrangian dual of the characterization of joint measurability. This leads to a simple criterion for joint measurability beyond the known qubit case.
\end{abstract}

\maketitle



{\it ``While [...] the wave function does not provide a complete description of physical reality, we left open the question of whether or not such a description exists. We believe, however, that such a theory is possible.}''\cite{EPR}

More than seventy years after Einstein, Podolsky and Rosen (EPR) raised this puzzle we know, as a consequence of Bell's argument \cite{Bell}, that a complete theory in the sense of EPR would force us to pay a high price--- such as giving up Einstein locality. Could there, however, be a theory which provides more information than quantum mechanics but still is `incomplete enough' to circumvent such fundamental conflicts? In this work we address a particular instance of this question, in the context of which the answer is clearly negative.

We consider observables which are not jointly measurable, i.e., \emph{incompatible} within quantum mechanics and show that they all enable the violation of a Bell inequality. That is, there exists a bipartite quantum state and a set of observables for an added site together with which the given observables violate a Bell inequality. As a consequence the observed probabilities do not admit a joint distribution \cite{Fine} unless this depends on the observable chosen at the added site, which conflicts with Einstein locality, i.e., the no-signaling condition (see appendix). So, if a hypothetical no-signaling theory is a refinement of quantum mechanics (but otherwise consistent with it \footnote{Here \emph{consistency} means that the hypothetical theory provides the same distributions as quantum mechanics; possibly after ignoring (integrating over) a hidden variable which would provide extra information.}), it can not render possible the joint measurability of observables which are incompatible within quantum mechanics---even if these observables are already almost jointly measurable in quantum theory.

An enormous amount of work has been done in related directions:  Bell inequalities \cite{Belloverview} and no-signaling theories \cite{nosignal} are lively fields of research. It is well known (and used in constructing quantum states admitting a local hidden variable description \cite{extension,extensions2}) that jointly measurable quantum observables can never lead to a violation of a Bell inequality \cite{Fine}. The converse, however, has hardly been addressed. For generalized measurements (POVMs) this might partly be due to the fact that no criterion for joint measurability is known beyond two-level systems, for which it was derived only recently \cite{twolevel}. A first indication of the present result can be found in \cite{Anderson} where it has been observed that for particular two-level observables the border of joint measurability \cite{BuschD} coincides with the one for the violation of the CHSH Bell inequality \cite{CHSH}. In the present work we show that the finding of \cite{Anderson} is not a mere coincidence resulting from having only few parameters, but that it holds in arbitrary dimension---even on a quantitative level. The main tool in this analysis will be the identification of the CHSH inequality with the Lagrangian dual of the joint measurability problem. This connection allows us at the same time to provide a simple criterion for joint measurability.
We will start, however, with a simpler case:

\section{Von Neumann measurements}

We begin with a warm-up on `sharp' observables, i.e., those described by Hermitian operators whose spectra represent the possible measurement outcomes. If a set of Hermitian operators is not simultaneously diagonalizable, then it contains at least one non-commuting pair. Similarly, such a pair of operators contains at least one non-commuting pair of spectral projections. By relabeling outcomes we can therefore always build a pair of non-commuting  $\pm 1$-valued observables $A_1$, $A_2$ from a set of incompatible von Neumann measurements. For each such pair we want to find now a bipartite quantum state and $\pm 1$-valued observables $B_1, B_2$ which violate the CHSH inequality $|\langle\mathbb{B}\rangle|\leq 1$ where
\be\label{eq:BCHSH}
\mathbb{B}=\frac12 \Big[A_1\otimes (B_1+B_2)+A_2\otimes (B_1-B_2)\Big].\ee
To this end note that for given observables the CHSH inequality holds for all quantum states iff $\mathbb{B}^2\leq\1$ \footnote{An inequality of the form $H_1\geq H_2$ between two Hermitian operators should be read as `$H_1-H_2$ is positive semidefinite'.}. Using that the observables have unit square  one gets \cite{Landau}
\be\label{eq:B2}
\mathbb{B}^2=\1+\frac14 [A_1,A_2]\otimes[B_1,B_2].\ee
Since the tensor product of the commutators is Hermitian and traceless, $\mathbb{B}^2$ has an eigenvalue larger than one iff the commutators do not vanish. Hence $B_i=A_i$ enables a violation whenever the observables $A_1$ and $A_2$ are incompatible.

For the optimal state (the respective eigenstate) this gives the quantitative relation
$|\avr{\mathbb{B}}|=\sqrt{1+||[A_1,A_2]||^2/4}$ whereas an optimal choice of the $B's$ ($||[B_1,B_2]||=2$, e.g. by fulfilling Pauli commutation relations) yields
\be\label{eq:Bmax1}\max_{\rho,B_1,B_2} \big|\avr{\mathbb{B}}\big|\;=\;\sqrt{1+\frac12\big|\big|[A_1,A_2]\big|\big|}\;.\ee

\section{General measurements}
Let $A_1$ and $A_2$ now be described by $d$-dimensional POVMs, i.e., pairs of positive semidefinite `effect' operators $\{Q,\1-Q\}$ and $\{P,\1-P\}$ whose expectation values give the probabilities of the assigned measurement outcomes. These observables are jointly measurable within quantum mechanics iff there is a measurement with four outcomes corresponding to four positive operators $R_{ij}$, $(i,j=\pm)$ with correct `marginals' $R_{++}+R_{+-}=Q$ and $R_{++}+R_{-+}=P$.

Beyond the case of qubits \cite{BuschD} there is no explicit characterization of jointly measurable observables known, but we can easily get an implicit one:
\begin{proposition}\label{prop:basic}
Two observables characterized by the effects $P$ and $Q$ are jointly measurable iff there is a positive semidefinite operator $S$ satisfying $Q+P-\1\leq S\leq P,Q$.
\end{proposition}
Necessity of this condition is proven by taking $S=R_{++}$ and sufficiency by simply constructing the other $R$'s from the given relations. A first look at Prop.\ref{prop:basic} suggests to just construct the `largest' $S$ which is smaller than $P$ and $Q$ and then check the two inequalities on the lower side. However, such a largest element does in general not exist (unless $P$ an $Q$ fulfill trivial relations such as $P\geq Q$ \cite{effects}) since, mathematically speaking, the set of positive operators does not form a lattice.

Despite this fact, Prop.\ref{prop:basic} can be decided efficiently numerically as it can be phrased as a \emph{semidefinite program} \cite{sdp} of the form
\be \inf \{\;\lambda\in\mathbb{R}\;|Q+P\leq \lambda\1+S\}\label{eq:sdp1}
\ee subject to the constraints $0\leq S\leq Q,P$.
The infimum of Eq.(\ref{eq:sdp1}), denote it $\lambda_0$, is larger than one iff the two observables are not jointly measurable. Moreover, the magnitude of $\lambda_0$ provides a means of quantifying how incompatible the two measurements are:  $\mu=\max[0,1-\lambda_0^{-1}]$ is the least amount of noise (or information loss) which has to be added to $Q,P$ in order to make the measurements $Q'=(1-\mu)Q+\mu E$ and $P'=(1-\mu)P+\mu E$ compatible for all $0\leq E\leq \1$.

Semidefinite programs always come with a dual (see Sec.\ref{sec:sdps}), and in case of (\ref{eq:sdp1}) deciding whether $\lambda_0>1$ (meaning that $Q$ and $P$ are incompatible) is equivalent to checking strict positivity of
\be\label{eq:actualdual} \lambda^*=\sup_{X,Y,Z\geq 0}\tr{X(Q+P-\1)}-\tr{QY}-\tr{PZ}\ee
 under the additional constraints $X\leq \rho$ where $\rho=Y+Z$ is a density operator.

Our aim is now to show that $\lambda^*$ is exactly the maximal violation of the CHSH inequality to which $Q$ and $P$ can lead and, using this insight, to provide a simple way of computing it (without the need of setting up a semidefinite programming algorithm).

\section{CHSH as Lagrangian dual}
Our main result is the following duality between the questions of whether two observables are jointly measurable and whether they enable a violation of the CHSH inequality:
\begin{proposition}[CHSH]
\label{prop:main}
Two measurements characterized by effect operators $Q$ and $P$ are not jointly measurable iff they enable the violation of the CHSH inequality. Quantitatively,
\be \sup_{\psi,B_1,B_2}\big|\langle\psi|\mathbb{B}|\psi\rangle\big|=1+2\lambda^*\label{eq:maineq}.\ee
 The supremum can be computed as  $\lambda^*=\max_{\phi\in[0,\pi]}\mu(\phi)$ where $\mu(\phi)$ is the largest eigenvalue of \be\label{eq:lambdaformula}
(Q+P-\1)\otimes\left(
  \begin{array}{cc}
    c^2 & cs \\
    cs & s^2 \\
  \end{array}
\right)- Q\otimes\left(
  \begin{array}{cc}
    1 & 0 \\
    0 & 0 \\
  \end{array}
\right)- P\otimes\left(
  \begin{array}{cc}
    0 & 0 \\
    0 & 1 \\
  \end{array}
\right), \ee with $c=\cos(\phi)$ and $s=sin(\phi)$.
\end{proposition}
\proof{We begin by rewriting the constraints in the  dual problem (\ref{eq:actualdual}) by introducing  $\rho:=Z+Y$, $\tilde{Q}:=\rho^{-1/2}X\rho^{-1/2}$ and $\tilde{P}:=\rho^{-1/2}Y\rho^{-1/2}$ (using the pseudo-inverse when necessary). The constraints in (\ref{eq:actualdual}) translate then to $0\leq \tilde{Q},\tilde{P}\leq \1$ (i.e., $\tilde{Q}$ and $\tilde{P}$ being effect operators) and $\rho$ being a density operator. We then exploit that the latter is the reduced density operator of a normalized pure state $|\psi\rangle:=(\sqrt{\rho}\otimes\1)\sum_{i=1}^d|ii\rangle$ and that for instance $\tr{QY}=\langle\psi|Q\otimes\tilde{P}^T|\psi\rangle$. In this way we obtain\bea
\lambda^*\hspace*{-5pt}&=&\hspace*{-3pt}\sup\;\langle\psi\big| \nonumber (Q+P-\1)\otimes\tilde{Q}-Q\otimes\tilde{P}-P\otimes(\1-\tilde{P})\big|\psi\rangle\\
&=&\hspace*{-3pt}\sup\;\langle\psi|\mathbb{B}-\1|\psi\rangle/2,\eea
where the supremum is taken over all admissible effect operators
$\tilde{Q},\tilde{P}$ and state vectors $\psi$ and the last step
is obtained by inserting $A_1=\1-2P$, $A_2=2Q-\1$,
$B_1=\1-2\tilde{P}$ and $B_2=\1-2\tilde{Q}$.

In order to arrive at the formula $\lambda^*=\max_{\phi\in[0,\pi]}\mu(\phi)$ we use that, due to convexity, the extremal value of $\avr{\mathbb{B}}$ is attained for $\tilde{P},\tilde{Q}$ being projections. Since two projections can be unitarily diagonalized simultaneously \cite{Halmos} up to blocks of size at most $2\times 2$, we obtain (again employing convexity) the same maximal violation when restricting to $\psi\in\mathbb{C}^d\otimes\mathbb{C}^2$. As the maximum over $\psi$ is nothing but computing the largest eigenvalue we can make further use of the unitary freedom we have to fix one of the observables, say $\tilde{P}={\rm diag}(1,0)$ and make $\tilde{Q}$ a real projector with non-negative entries, which finally leads to the expression in (\ref{eq:lambdaformula}).
}

\section{Discussion}
The two cases discussed above, von Neumann measurements and POVMs, differ in several respects:
we saw in the von Neumann case that it is sufficient for a CHSH-violation to use the same observables at the added site. In the case of POVMs this is no longer true. To see this, note that a rescaling $A_i\mapsto \lambda A_i$ implies $\langle \mathbb{B}\rangle\mapsto\lambda \langle \mathbb{B}\rangle$. This means that starting with von Neumann measurements on both sides and rescaling the $A_i$'s until right before the violation vanishes will not allow us anymore to depart from von Neumann observables for the $B_i$'s.

A second difference is the reduction argument which allowed us for von Neumann measurements to reduce the case of many observables with several outcomes to two $\pm1$-valued observables. The importance of this step stems from the fact that the `two $\pm1$-valued observable case' is the only one where a single Bell inequality is sufficient to characterize the existence of a joint probability distribution\cite{FineCHSH}. For more observables or outcomes the set of Bell inequalities becomes fairly monstrous and is largely unexplored. Unfortunately, a similar reduction to the CHSH-case is not always possible for POVMs: incompatibility of observables can in this case be `overlooked' if one only considers pairwise incompatibility of several observables or if one groups measurement outcomes together \cite{Teiko}.
Hence, for the cases of POVMs (with more than two outcomes or settings) where this happens the question is still open. If one follows the same route as in the proof of Prop.\ref{prop:main} one easily arrives at expressions which show that the given observables can be used as one-side part of an \emph{entanglement witness}. However, not every witness corresponds to a Bell inequality and whether or not this is the case highly depends on the type of decomposition into local operators. We have to leave this problem open for the moment.

Finally, it is worth mentioning that the followed approach led to new insight into the joint measurability problem. On the one hand Prop.(\ref{prop:main}) provides a simple criterion for deciding joint measurability for two two-valued observables beyond the recently proven qubit case \cite{twolevel}. On the other hand, the fact  that the problem is a semidefinite program enables us  to solve it in practice (i.e., for any given instance). As this is an interesting result in its own right we will provide more details about it in the remaining part of this paper.

\section{Joint measurability as a semidefinite program\label{sec:sdps}}
The fact that the joint measurability problem is a semidefinite program implies that for every instance of observables (with not too many parameters) the problem can be solved in an efficient and certifiable way. As we saw that the dual problem is related to the violation of a Bell inequality we will state the problems  in a quantitative way (i.e., not as mere feasibility problem).

The duality theorem for semidefinite programs \cite{sdp} reads
\bea\label{eq:eqsdpdThm1}&& \inf_{x\in\mathbb{R}^n}\left\{\langle c|x\rangle\; \big|\; \sum_i x_i F_i\geq C\right\}\\
&\geq& \sup_{X\geq 0}\left\{\tr{C X}\; \big|\; \tr{X F_i}=c_i\right\}\label{eq:eqsdpdThm2},\eea
where $c\in\mathbb{R}^n$ and $C,F_i$ are Hermitian matrices. Moreover, if one of the problems (say the primal problem (\ref{eq:eqsdpdThm1})) is bounded and strictly feasible (i.e.,  $\exists x:\sum_ix_iF_i>C$), then the dual attains its extremum and equality holds between (\ref{eq:eqsdpdThm1}) and (\ref{eq:eqsdpdThm2}).

\subparagraph{Two dichotomic observables} The joint measurability problem (\ref{eq:sdp1}) can be cast as a semidefinite program (\ref{eq:eqsdpdThm1}) by expanding $S=\sum_i x_i G_i$ in terms of a Hermitian operator basis $\{G_i\}$ and setting \bea C&=& (Q+P)\oplus 0\oplus (-Q)\oplus (-P),\\
F_0 &=& \1\oplus 0\oplus 0\oplus 0,\\
F_i &=& G_i\oplus G_i\oplus (-G_i)\oplus (-G_i),\ i\geq 1\eea and $c_0=1$, $x_0=\lambda$, $c_i=0$ for $i\geq 1$.
From here we get the dual
\be\label{eq:dual1} \sup_{\rho, Y,Z} \tr{\rho(Q+P)}-\tr{QY}-\tr{PZ},\ee
subject to the constraints $Y,Z\geq 0$ and $\rho\leq Y+Z$ being a density operator. As this is strictly feasible, the supremum in (\ref{eq:dual1}) coincides with the minimum $\lambda_0$ of (\ref{eq:sdp1}). For our purposes we slightly rewrite the problem and instead of checking whether $\lambda_0> 1$ (which means that $Q$ and $P$ are not jointly measurable) we may as well study whether
\be \sup_{X,Y,Z\geq 0}\tr{X(Q+P-\1)}-\tr{QY}-\tr{PZ}\ee
is positive under the constraint $X\leq Y+Z$ and $\tr{Y+Z}=1$. This is the form used in (\ref{eq:actualdual}). The corresponding primal problem changes the constraints $S\leq P,Q$ and $P+Q\le \lambda\1+S$ in Prop.\ref{prop:basic} to $S-\lambda\1\leq P,Q$ and $P+Q\le S+\1$, and minimizes $\lambda$ leading to the minimum $\lambda^*$.

In a similar vein we can now treat more general scenarios. All of them have a strictly feasible dual so that equality holds between primal and dual problem.

\subparagraph{Two arbitrary observables} Consider two $N-$outcome observables which are characterized by two sets of effect operators $\{Q_i\}$,$\{P_j\}$ with $i,j=1,\ldots, N$. These are jointly measurable iff we can find $\{R_{ij}\geq 0\}$ such that $\sum_iR_{ij}=P_j$ and $\sum_jR_{ij}=Q_i$. One way, analogous to the previous one, to express this as a semidefinite program is to minimize $\lambda\in\mathbb{R}$ w.r.t. to $\{R_{ij}\geq 0\}$ such that
\be\sum_{i=1}^{N-1}Q_i+\sum_{j=1}^{N-1}P_j\leq \lambda\1+\sum_{i,j=1}^{N-1}R_{ij}\ee
and  $P_j\geq\sum_{i=1}^{N-1}R_{ij}$ and $Q_i\geq\sum_{j=1}^{N-1}R_{ij}$ for all $i,j$.
The corresponding dual is
\be \lambda_0=\sup_{\rho,\{Y_i,Z_j\}\geq 0} \sum_{i=1}^{N-1}\tr{Q_i(\rho-Y_i)+P_i(\rho-Z_i)},\ee
subject to the additional constraints $\rho\leq Y_i+Z_j$ for all $i,j$ and $\rho$ being a density operator.

\subparagraph{Several dichotomic observables} Let $M$ two-valued observables be characterized by  effect operator $0\leq T_\alpha\leq\1$, $\alpha=1,\ldots, M$.
We will denote the effect operators of the sought joint observable by $R_i$ using a multi-index $i\in\{0,1\}^M$ with $|i|:=\sum_\alpha i_\alpha$. $T_\alpha$ will be identified with the sum of all $R_i$ for which $i_\alpha=1$. The existence of a joint observable can then be expressed in terms of the constraints $R_i\geq 0$ and
\bea \forall\alpha:\quad\sum_{|i|>1}R_i\delta_{i_\alpha,1}&\leq& T_\alpha,\\
\sum_\alpha T_\alpha &\leq& \1+\sum_{|i|\geq 1}(|i|-1)R_i.\eea
This is again a semidefinite program which can be made quantitative by replacing $\1\rightarrow\lambda\1$ and minimizing $\lambda$. Again the minimum $\lambda_0$ can as well be  obtained from the dual
\bea \lambda_0 &=& \sup_{\rho,\{X_\alpha\geq 0\}} \sum_\alpha \tr{T_\alpha(\rho-X_\alpha)}, \\
&&\mbox{subject to}\ \ \nonumber\forall i:\ (|i|-1)\rho\leq\sum_{\alpha} \delta_{i_\alpha,1} X_\alpha, \eea
where $\rho$ is constrained to be a density operator.

\section{appendix: Bell inequalities and no-signaling}

For completeness we provide in this appendix the argument for the claim that observables which enable the violation of a Bell inequality cannot be measured jointly within any no-signaling theory which is consistent with the predictions of quantum mechanics (but possibly a refinement thereof). Variants of this argument (or its main ingredients) can be found in \cite{Fine,nosignal,extensions2,WernerInv}.

Suppose Alice can jointly measure two observables, which are labeled by $A_1$ and $A_2$, and yield outcomes $a_1$, $a_2$ with probability $p(a_1,a_2)$. If Bob, at a distance, measures an observable $B_1$ with outcome $b_1$, then they observe in a statistical experiment a joint probability distribution $p(a_1,a_2,b_1|B_1)$ so that \footnote{As common in classical probability theory we use the notation $p(\cdot|\cdot)$ where right of the dash is the \emph{condition} which the probability is subject to. In our case this is the observable which is measured. }
\be
 p(a_1,a_2)=\sum_{b_1} p(a_1,a_2,b_1|B_1).
\ee
However, in a no-signaling theory this has to be independent of Bob's chosen observable, i.e., a possibly measured $ p(a_1,a_2,b_2|B_2)$ has to have the same marginal $p(a_1,a_2)$. Assume that Bob chooses $B_1$ or $B_2$ at random so that they measure both triple distributions. From these we can write down a joint distribution
\be \label{eq:jointcl}
p(a_1,a_2,b_1,b_2):=\frac{p(a_1,a_2,b_1|B_1)p(a_1,a_2,b_2|B_2)}{p(a_1,a_2)},\ee
which by construction correctly returns all measured distributions as marginals.
As a result, the possibility of jointly measuring $A_1$ and $A_2$ implies a joint probability distribution (\ref{eq:jointcl}) if the no-signaling condition is invoked. A joint distribution, in turn, implies that no Bell inequality can be violated. So if a Bell inequality is violated, then either $A_1$ and $A_2$ are not jointly measurable, or the no-signaling condition is violated.

Note that this argument works independent of the numbers of measurement outcomes or observables.

\ 

This work has been funded by Spanish grants FPU, I-MATH and MTM2008-01366, by QUANTOP and the Ole Roemer grant of the Danish Natural Science Research Council (FNU). MMW thanks UCM/ISI for the hospitality.

\end{document}